\documentclass{JHEP3} 

\usepackage{epsfig,multicol}

\newcommand\fverb{\setbox\pippobox=\hbox\bgroup\verb}
\newcommand\fverbdo{\egroup\medskip\noindent%
			\fbox{\unhbox\pippobox}\ }
\newcommand\fverbit{\egroup\item[\fbox{\unhbox\pippobox}]}
\newbox\pippobox

\title{A simple method to simulate $W/Z$ + jet production at hadron collisions 
with PS-ME matching}

\author{Shigeru Odaka\\
	High Energy Accelerator Research Organization (KEK)\\
        1-1 Oho, Tsukuba, Ibaraki 305-0801, Japan\\
	E-mail: \email{shigeru.odaka@kek.jp}}

\abstract{
We propose a simple method to simulate $W/Z$ + jet productions at hadron 
collisions.
The simulation can be done by using existing tools with some modifications 
allowed to users.
$W/Z$ + 1 jet events are generated using an ME-based event generator 
at the tree level.
The divergence at low $p_{T}$ is suppressed by using the Sudakov form factor.
PS is added with an appropriate consideration for PS-ME matching.
The simulation for the $W$ + 1 jet production shows a smooth matching 
with the $W$ + 0 jet simulation at low $p_{T}$s.
The $Z$ + 1 jet simulation in the Tevatron Run I condition well reproduces 
the experimental measurement on the $p_{T}$ spectrum of $Z$.
}


\begin{document} 


\section{Introduction}
The simulation of multiple hadron jet (multi-jet) production is one of 
the most serious problems in physics analyses 
at high-energy hadron collider experiments.
Our experimental reach extends to heavier objects as the collision energy 
increases.
Once such heavy objects decay hadronically, they frequently produce multi-jet 
final states.
Similar multi-jet configuration originating from well-separated emission of 
light quarks and/or gluons can be easily produced by non-resonant 
QCD interactions.
They may blind or distort interesting heavy object signals.
We have already encountered this problem in the study of hadronic top-quark 
decays at Tevatron.
The problem will become much more severe for heavier objects 
(Higgs boson(s), SUSY particles, {\it etc.}) expected at forthcoming LHC 
experiments, 
since in principle the signal frequency decreases as the cm energy of the 
hard interaction increases while the QCD activity remains nearly constant.
Understanding of the QCD multi-jet production will become crucial for 
the study of such heavy objects.
Despite that, the technique is not well established for the simulation.

The simulation of hadron collision interactions consists of two 
components: parton showers (PS) describing relatively soft regions in the 
initial-state and the final-state interactions, 
and a simulation of hard interactions based on perturbative 
matrix-element (ME) calculations.
The former is a 3-dimensional model based on the factorization theory, 
adding the contributions of large collinear components to all orders.
The matching between the two components is important, but is not trivial 
since theoretical discussions are usually made only at the collinear limit 
where transverse behaviors are ignored.
A certain model-based discussion is necessary to construct a 
consistent multi-jet simulation.

Recently, the CKKW technique originally proposed for the final-state 
jet production at $e^{+}e^{-}$ collisions \cite{ckkw} has been extended to the 
initial state \cite{ckkw_had}, and implemented in an event generator 
for hadron collisions \cite{sherpa}.
This technique consistently adds tree-level ME calculations 
for a certain hard interaction process associated with 0 jet, 1 jet, 2 jets, 
..., to provide an exclusive event sample above a certain resolution scale.
A PS is applied to a limited phase space which ME calculations do not cover, 
with a careful consideration for the matching.
The concept is also implemented in different ways in other event generators 
\cite{other_gen}.

CKKW provides us with an exclusive multi-jet sample.
Despite, in many cases, users are interested only in an inclusive behavior 
of a certain number of leading jets.
They ignore non-leading ones in order to make the evaluation 
as free from theoretical and experimental ambiguities in soft regions 
as possible.
We expect that there must be a simpler method for such applications.
In this paper we propose a method to obtain a simulation sample 
of inclusive $n$-jet events by using an $n$-jet ME event generator 
with the help of an appropriate PS simulation.

In the present study 
we focus on the simplest case, the $W$ + 1 jet production.
Namely, we are interested only in the leading-jet behavior produced in 
association with the $W$ boson production.
This process deals with the jet emission from the initial-state partons only.
The discussions would be able to extend to the final state with appropriate 
replacements of the parameters and formulae.
We use only those tools which are publicly available.
Modifications are applied at the level where ordinary users are allowed.
We apply our method to the $Z$ + 1 jet production in the Tevatron Run I 
condition for a comparison with experimental measurements. 
The $Z$ production is used only because the momentum measurement is expected 
to be less ambiguous than the $W$ production.

\section{$W + 1$ jet simulation}

The $W + 1$ jet production is simulated at the tree level 
by using GR@PPA version 2.76 \cite{grappa27} together with the PS 
in PYTHIA version 6.212 \cite{pythia62} in the present study.
The $W$ bosons always decay to the pair of an electron and a neutrino.
GR@PPA includes this decay in the ME calculation.
The sample program in the GR@PPA distribution package is used for 
interfacing GR@PPA and PYTHIA.
Studies are done for the LHC condition, proton-proton collisions 
at the cm energy of 14 TeV.
A CTEQ6L1 \cite{cteq6l1} routine included in the sample program is used for PDF.

Both the initial and final state PSs are activated in PYTHIA with the default 
setting, while the hadronization and the multiple interaction are deactivated 
for simplicity ({\tt MSTP(81) = 0} and {\tt MSTP(111) = 0}).
The QED radiation is also deactivated ({\tt MSTJ(41) = 1}).
A jet clustering (PYCELL in PYTHIA) is applied to the parton-level events,
where the detector is assumed to cover the full azimuth ($\phi$) 
and the pseudorapidity ($\eta$) up to 4.5 in the absolute value 
with a granularity of about 0.1 in both $\phi$ and $\eta$. 
The jets are reconstructed using a cone algorithm with the half-cone size of 
0.4 in $R$ and with the transverse energy ($E_{T}$) threshold of 10 GeV.

A naive simulation of this process has an obvious difficulty.
In hadron collision simulations we have to give four energy scales: 
renormalization scale ($\mu_{R}$), factorization scale ($\mu_{F}$), 
and energy scales for the initial-state PS ($\mu_{\rm ISR}$) 
and the final-state PS ($\mu_{\rm FSR}$).
It is usually said that they should be set to the {\it typical} energy scale 
of the interaction.
However, there are two energy scales in the $W + 1$ jet production: 
the $W$-boson mass ($m_{W}$) and the transverse momentum ($p_{T}$) of the jet.
They can be quite different from each other if we allow very low $p_{T}$ 
($\sim$ 10 GeV/$c$) for the jet.

\EPSFIGURE[t]{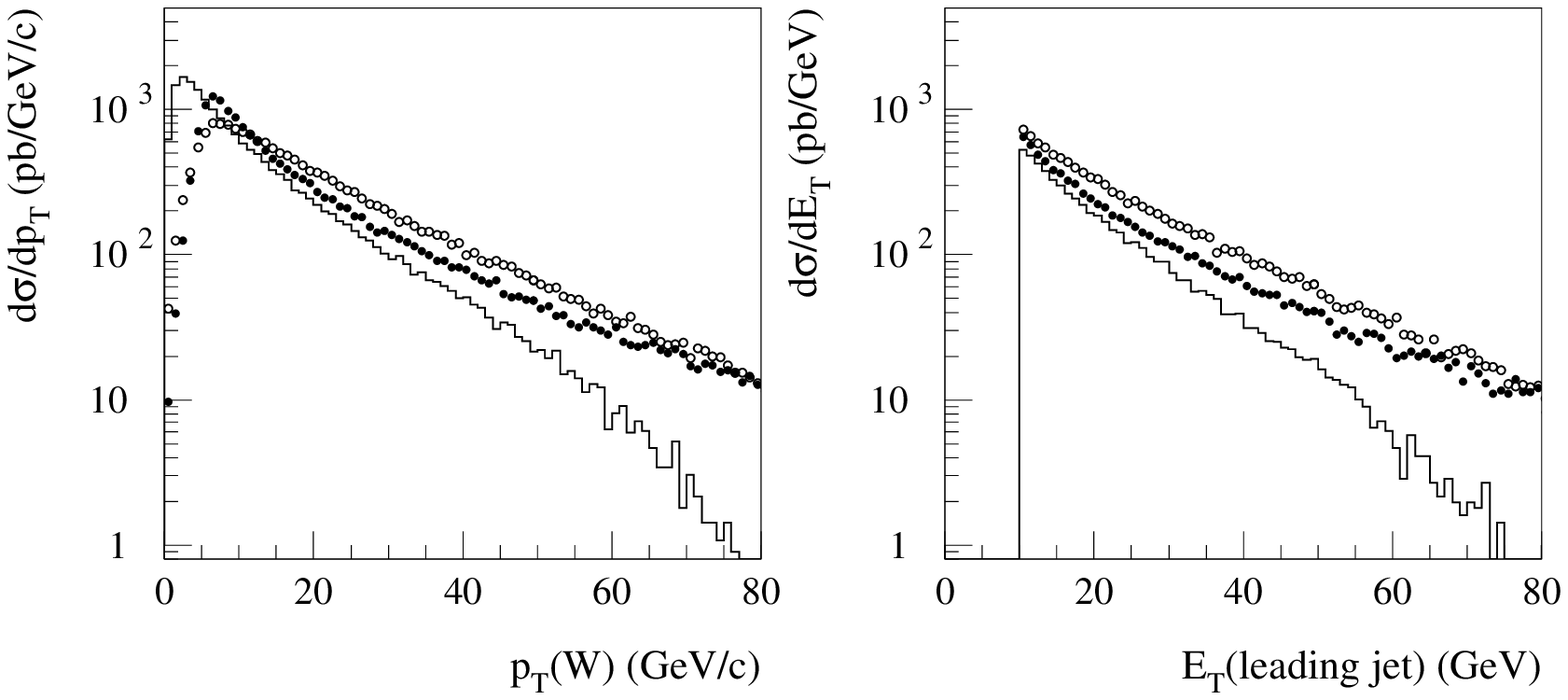,width=160mm}
{\label{fig_simple}
Naive simulation of $W$ + 1 jet production at LHC.
The $p_{T}$ distribution of the $W$ boson and the $E_{T}$ distribution of 
the leading jet are plotted.
The $W$ + 1 jet events are generated by GR@PPA 2.76 with the ${\hat p}_{T}$ 
cut of 5 GeV/$c$, and the PYTHIA PS is applied to the initial and final states 
with the default setting.
A $E_{T}$ cut of 10 GeV is imposed in the jet reconstruction.
Results for two extreme cases of the energy-scale choice are shown: 
$\mu = {\hat p}_{T}$ with solid circles, and $\mu = m_{W}$ with open circles.
Histograms show the results from the $W$ + 0 jet simulation with $\mu = m_{W}$.
}

Figure \ref{fig_simple} shows the simulation results for the $p_{T}$ 
distribution of $W$ and the $E_{T}$ distribution of the leading jet 
(the highest $E_{T}$ jet).
The ME events are generated with a minimum $p_{T}$ cut of 5 GeV/$c$ 
in the cm frame of the hard interaction.
Distributions are shown for two extreme cases of the energy-scale choice.
All energy scales are defined to be identical 
and set equal to the jet $p_{T}$ in the cm frame (${\hat p}_{T}$) in one case, 
and the $W$ mass (the invariant mass of the electron and neutrino) in the other.
These choices are non-standard in GR@PPA.
The user-define option ({\tt ICOUP = IFACT = 6}) is selected, 
and the energy-scale parameters ({\tt GRCQ} and {\tt GRCFAQ}) are appropriately 
set in the subroutine GRCUSRSETQ.
We frequently see similar simulations in previous $W$ + jets analyses.
The figure shows that the difference is significant not only in the absolute 
value but also in the shape.

The prediction from the $W$ + 0 jet event generator in GR@PPA 2.76 
is also shown in Fig. \ref{fig_simple}.
The same PYTHIA PS is applied with all the energy scales set equal 
to the $W$ mass.
The finite $p_{T}$ of $W$ and all jet activities are generated 
by PS in this case.
The low $p_{T}$ behavior must be well described by this simulation.
However, neither the two $W + 1$ jet simulations agrees with it at low $p_{T}$s.
The $\mu = {\hat p}_{T}$ simulation looks better around $p_{T}$ = 20 GeV/$c$, 
but deviates at further low $p_{T}$s.
This is unavoidable since the $W + 1$ jet cross section diverges 
at ${\hat p}_{T} = 0$. 


\begin{figure}[t]
  \centering
  \epsfig{file=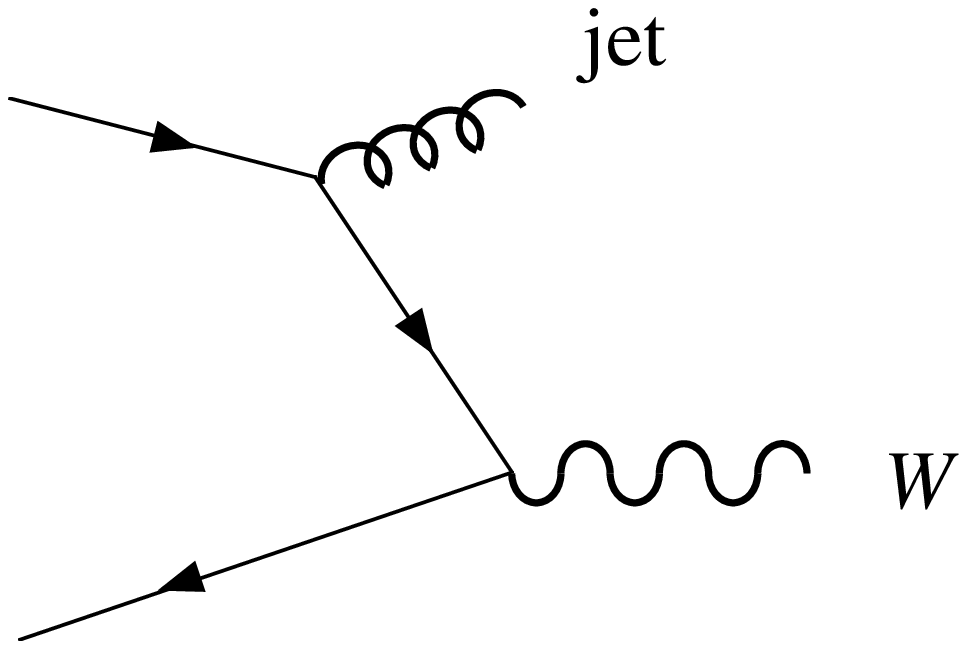,width=70mm}
  \caption{One of the Feynman diagrams for which the $W$ + 1 jet ME 
is evaluated.
  \label{fig_diagram}}
\end{figure}

Figure \ref{fig_diagram} illustrates one of the Feynman diagrams for which 
the $W + 1$ jet ME is evaluated.
In the present {\it inclusive} analysis, the jet in the ME should be 
identical to the leading jet that we observe.
This means that there should not be any jet having $p_{T}$ larger than that.
The naive simulation with $\mu = m_{W}$ is apparently inadequate from this 
point of view.
The initial-state PS frequently generates high $p_{T}$ jets, 
resulting in large cross sections in medium $p_{T}$ and $E_{T}$ regions 
in Fig. \ref{fig_simple}.

The excessive PS activities can be eliminated by the choice of 
$\mu = {\hat p}_{T}$.
This leads to a better behavior in low $p_{T}$ and $E_{T}$ regions.
The divergence at further low $p_{T}$s appears only because we stop 
the perturbation at a limited order, the lowest order in this case.
Higher order corrections would produce higher $p_{T}$ jets and suppress 
the contribution of the $W$ + 1 jet ME at low $p_{T}$s.
This is considered to be the mechanism to make the actual observation finite.
The suppression can be evaluated to all orders of the coupling constant 
in a collinear approximation in terms of the Sudakov form factor.
This is the way usually adopted in QCD calculations having multiple energy 
scales, and the way used in PS.
Therefore, the application of such a suppression with an appropriately 
evaluated Sudakov form factor should modify the $W + 1$ jet simulation 
to match the $W$ + 0 jet simulation at low $p_{T}$s, and to be finite 
even at $p_{T}$ = 0.
We look for a suitable expression of the Sudakov form factor in the next 
section.

Although the choice of $\mu = {\hat p}_{T}$ looks better, it is too naive 
to construct a realistic simulation.
We reconsider the choice of energy scales and discuss about the matching 
between PS and ME in later sections.

\section{Sudakov form factor}

The Sudakov form factor for quarks can be described as 
\begin{equation}\label{sudakov1}
S_{q}(Q_{1}^{2}, Q_{2}^{2}) = \exp\left[ - \int_{Q_{1}^{2}}^{Q_{2}^{2}}
{dQ^{2} \over Q^{2}} \int_{0}^{1-\epsilon} dz \  
{\alpha_{s} \over 2\pi}\ P_{q \rightarrow q}(z) \right] ,
\end{equation}
where the splitting function for quarks with the radiation of a timelike 
gluon is 
\begin{equation}\label{split}
P_{q \rightarrow q}(z) = C_{F} { 1 + z^{2} \over 1 - z}
\end{equation}
at the leading order, with $C_{F} = 4/3$.
Equation (\ref{sudakov1}) represents the probability that a quark or an 
anti-quark survives without any radiation from an evolution parameter 
value of $Q_{1}^{2}$ up to $Q_{2}^{2}$.
However, the quantity is unphysical since the result depends on an 
artificial cutoff $\epsilon$.
A smaller $\epsilon$ gives a smaller $S_{q}$.
This corresponds to a natural explanation that the non-radiation probability 
becomes smaller if we allow softer radiation, 
but the softness is not well defined.
It is necessary to introduce a definition of the splitting kinematics 
in order to derive a physically meaningful quantity.

\EPSFIGURE[t]{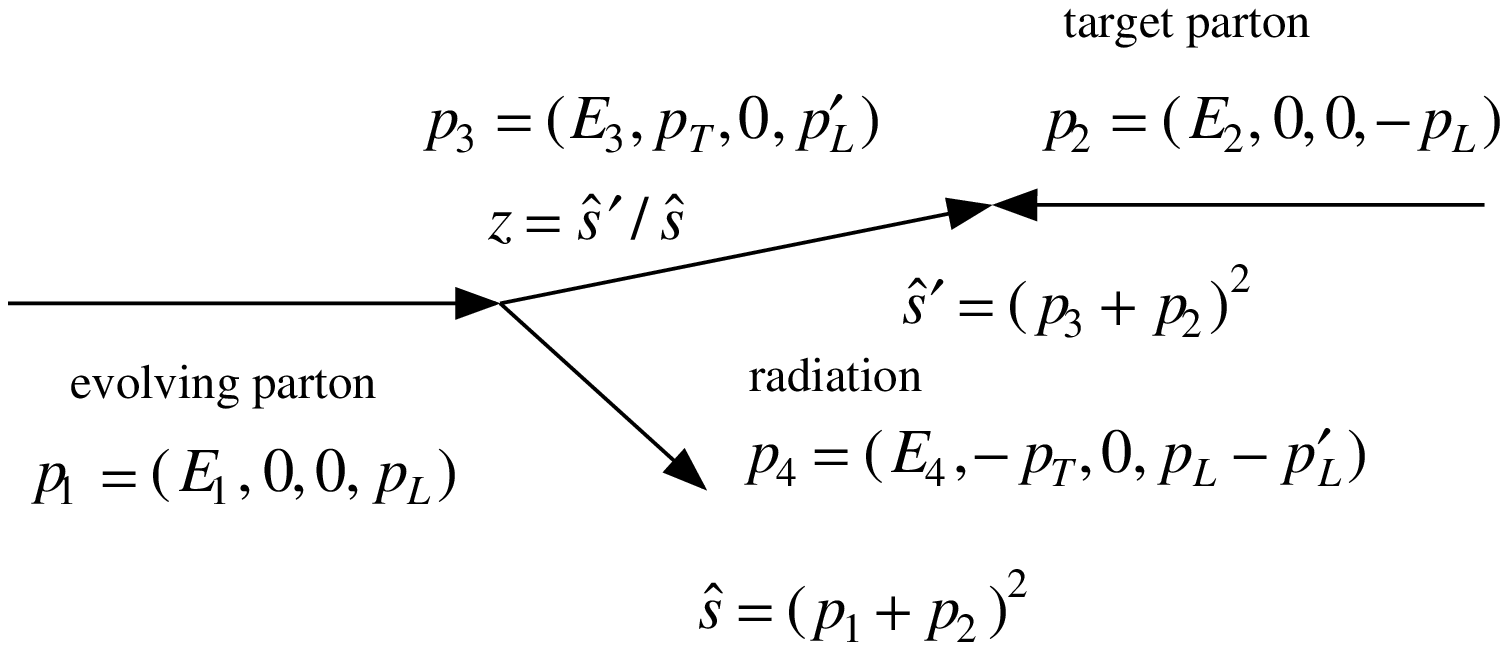,width=100mm}
{\label{fig_kinem}
The definition of the PS splitting kinematics that we adopt.}

Here we examine the definition in PYTHIA for the initial-state PS 
\cite{pythia_kinem},
where $Q^{2}$ is exactly the virtual of evolving partons.
Details of the definition are shown in Fig. \ref{fig_kinem}.
The four-momenta are so defined that $p_{1}^{2}$, $p_{2}^{2}$ and 
$p_{3}^{2} < 0$, and $p_{4}^{2} \geq 0$.
The splitting parameter $z$ is the ratio of the squared invariant mass of 
the collision system after and before the splitting, given as
\begin{equation}
z = { \hat{s}^{\prime} \over \hat{s} } 
= { (p_{3} + p_{2})^{2} \over (p_{1} + p_{2})^{2} } .
\end{equation}
This definition preserves the relation $\hat{s}_{\rm hard} = x_{1}x_{2}s$
with $x$ given by the product of all $z$ values in each beam.
The squared four-momenta, $\hat{s}$ and $z$ are the inputs.
The other parameters can be derived from them.

In this definition, the transverse momentum of the splitting can be 
described as \cite{pythia_kinem}
\begin{equation}
p_{T}^{2} = E_{3}^{2} - p_{L}^{\prime 2} - p_{3}^{2},
\end{equation}
where
\begin{eqnarray}
E_{3} = { \hat{s}^{\prime} + p_{1}^{2} - p_{2}^{2} - p_{4}^{2} 
\over 2\sqrt{\hat{s}} } , \qquad
p_{L}^{\prime} = { \hat{s}^{\prime} - p_{2}^{2} - p_{3}^{2} - 2E_{2}E_{3} 
\over 2p_{L}} , \nonumber
\end{eqnarray}
with
\begin{eqnarray}
p_{L}^{2} = { (\hat{s} - p_{1}^{2} - p_{2}^{2})^{2} - 4 p_{1}^{2} p_{2}^{2}
\over 4\hat{s} } , \qquad
E_{2} = { \hat{s} - p_{1}^{2} + p_{2}^{2} 
\over 2\sqrt{\hat{s}} } . \nonumber
\end{eqnarray}
In order to simplify the relation 
we assume the radiation to be massless ($p_{4}^{2} = 0$), 
and take the limit $p_{2}^{2} \rightarrow 0$ since fundamental properties 
should not depend on not-well-defined target virtuality.
Furthermore, we take the limit $p_{3}^{2} \rightarrow p_{1}^{2} = -Q^{2}$ 
since Eq. (\ref{sudakov1}) indicates that we can take the 
virtuality step as small as we want.
These approximations lead us to a relation, 
\begin{equation}\label{relation1}
p_{T}^{2} = { ( 1 - z )^{2} Q^{2} \over 1 + { Q^{2} \over \hat{s} } } .
\end{equation}
If we assume $Q^{2} \ll \hat{s}$, this can be further simplified to 
\begin{equation}\label{relation2}
p_{T} = ( 1 - z ) Q .
\end{equation}
This is different from the usually quoted relation 
$p_{T}^{2} = ( 1 - z ) Q^{2}$, 
but similar to the one obtained for the HERWIG PS \cite{herwig_ps}.


\begin{figure}[t]
  \centering
  \epsfig{file=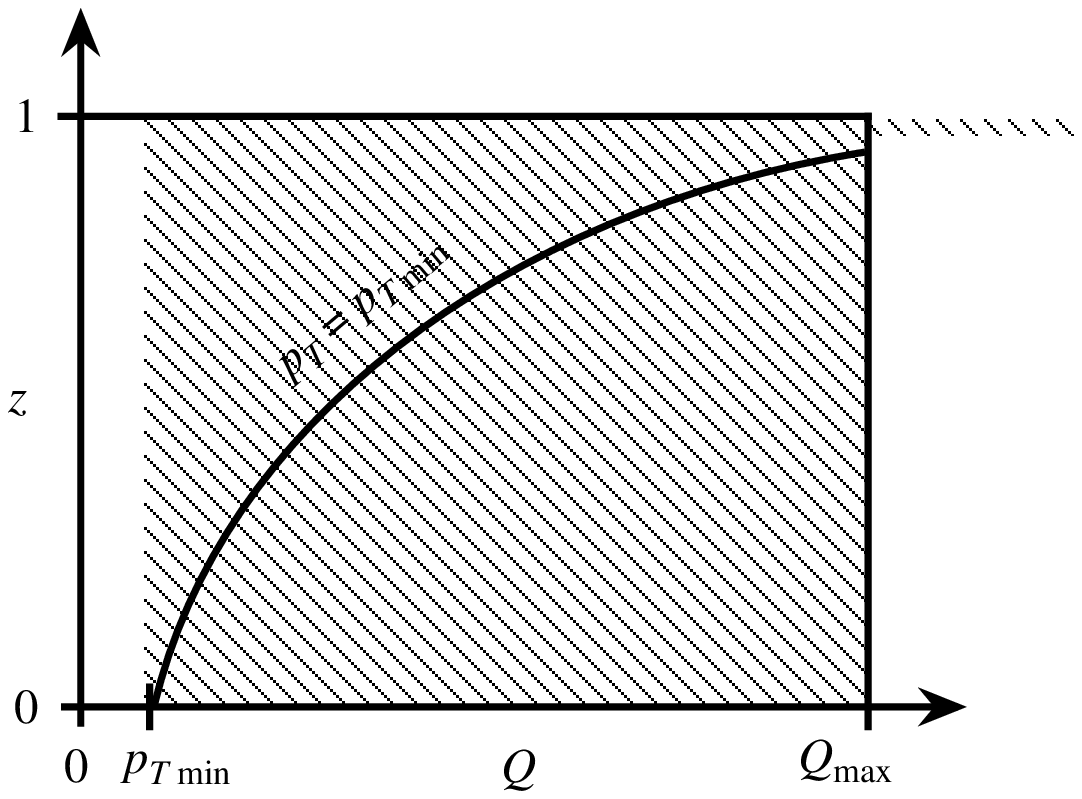,width=70mm}
  \caption{The integration area defined by the $\theta$ function 
in Eq. (\protect\ref{sudakov2}).}
  \label{fig_integ}
\end{figure}

If we assume the relation (\ref{relation2}), the Sudakov form factor 
giving the no-radiation probability above a certain $p_{T}$ cut 
($p_{T{\rm min}}$) can be calculated as
\begin{equation}\label{sudakov2}
S_{q}(p_{T{\rm min}}, Q_{\rm max}) = \exp\left[ - \int_{0}^{Q_{\rm max}^{2}}
{dQ^{2} \over Q^{2}} \int_{0}^{1} dz \  
{\alpha_{s} \over 2\pi}\ P_{q \rightarrow q}(z) 
\theta(p_{T} - p_{T{\rm min}}) \right] .
\end{equation}
The integration area defined by the $\theta$ function is schematically 
illustrated in Fig. \ref{fig_integ}.
The $\theta$ function naturally gives an upper bound of the $z$ integration 
and an lower bound of the $Q^{2}$ integration.
Therefore, there is no need to introduce any artificial cutoff.
Namely, the result is well-defined and must have a physical meaning.
The upper bound $Q_{\rm max}$ should be given by the hardest energy scale in 
the considered interaction.
It would be natural to take the $W$-boson mass for non-hard radiation 
in the $W + 1$ jet production.
Although Eq. (\ref{sudakov2}) is numerically calculable, 
it is rather complicated for the application to actual simulations.

Let's start from the radiation function which the Sudakov form factor is 
based on.
The radiation probability is usually expressed as
\begin{equation}
d\Gamma = {dQ^{2} \over Q^{2} }dz\ {\alpha_{s} \over 2\pi}\ P(z) .
\end{equation}
If we assume the relation (\ref{relation2}), we obtain another expression 
that 
\begin{equation}
d\Gamma = {dp_{T} \over p_{T} }dz\ {\alpha_{s} \over \pi}\ P(z) .
\end{equation}
Using this expression, the Sudakov form factor having the same meaning 
as Eq. (\ref{sudakov2}) can be written as
\begin{equation}\label{sudakov3}
S_{q}(p_{T{\rm min}}, Q_{\rm max}) = \exp\left[ 
- \int_{p_{T{\rm min}}}^{Q_{\rm max}}
{dp_{T} \over p_{T}} \int_{0}^{1-{p_{T} \over Q_{\rm max}}} dz \  
{\alpha_{s} \over \pi}\ P_{q \rightarrow q}(z) \right] .
\end{equation}
The upper bound of the $z$ integration is given by the relation 
$p_{T} = (1 - z)Q \leq (1 - z)Q_{\rm max}$.

Since it is natural to define $\alpha_{s}$ to be a function of $p_{T}$ 
of the splitting, the $z$ integration is easy to perform.
Assuming the leading order function (\ref{split}), we obtain
\begin{equation}\label{sudakov4}
S_{q}(p_{T{\rm min}}, Q_{\rm max}) = \exp\left[ 
- \int_{p_{T{\rm min}}}^{Q_{\rm max}}
{2C_{F} \over \pi}\ {\alpha_{s}(p_{T}) \over p_{T}}\ 
\left( \ln{1 \over \epsilon} - {3 - 4\epsilon + \epsilon^{2} \over 4} 
\right) \ dp_{T} \right]
\end{equation}
with $\epsilon = p_{T}/Q_{\rm max}$.
This leads to the expression used in the CKKW method \cite{ckkw} if we take 
the limit $\epsilon \rightarrow 0$ in the non-divergent term, 
although the definition of the parameters is slightly different.
The difference between Eq. (\ref{sudakov2}) and Eq. (\ref{sudakov4}) is 
only in the definition of the integration variables.
They give exactly identical answers.

\section{Suppressed $W + 1$ jet simulation}

For the test of the suppression, $W$ + 1 jet events are generated 
using GR@PPA 2.76 with the energy scale choice of 
$\mu_{R} = \mu_{F} = {\hat p}_{T}$,
where ${\hat p}_{T}$ is the transverse momentum 
of the jet and $W$ in the cm frame of the hard interaction.
The choice $\mu_{R} = {\hat p}_{T}$ merely means that the coupling at the 
parton splitting is evaluated at this energy scale.
A CKKW-like correction has to be applied when we have multiple QCD vertices.
The other conditions are the same as those in the simulations 
leading to Fig. \ref{fig_simple}.

The weight factor for the suppression is defined as 
\begin{equation}
w = S_{q}({\hat p}_{T}, Q_{\rm max})^{2}
\end{equation}
with Eq. (\ref{sudakov4}),
where we use the first-order expression for the strong coupling: 
\begin{equation}
\alpha_{s}(p_{T}) = { 4\pi \over \beta_{0}\ln({p_{T}^{2}/\Lambda^{2}) } }
\end{equation}
with $\beta_{0} = 11 - 2n_{f}/3$ ($n_{f} = 5$) and $\Lambda$ = 0.0883 GeV.
This gives $\alpha_{s}(m_{Z})$ = 0.118.
The hard interaction scale is given by
\begin{equation}
Q_{\rm max} = {\hat m}_{T}(W) = \sqrt{ m_{W}^{2} + {\hat p}_{T}^{2} }.
\end{equation}
This naturally connects reasonable definitions at two extreme cases: 
$Q_{\rm max} = m_{W}$ at small ${\hat p}_{T}$s and $Q_{\rm max} = 
{\hat p}_{T}$ at large ${\hat p}_{T}$s.
The Sudakov form factor is squared since the no-radiation condition must be
required to two incoming quarks converted to the $W$ boson.
Events generated according to the ordinary $W$ + 1 jet cross section 
are accepted in proportion to this weight 
in the subroutine GRCUSRCUT of GR@PPA.
Though this degrades the event generation efficiency, 
the degradation is moderate in GR@PPA because the same routine is used 
in the initialization stage for optimizing the event generation.

We have to apply PS also to these $W$ + 1 jet events in order to obtain 
realistic events.
Here, an appropriate care is necessary to accomplish a reasonable matching 
between the PS and ME.
In principle the PYTHIA PS is equivalent to the numerical evaluation of 
the integral in Eq. (\ref{sudakov1}).
The Sudakov form factor is evaluated by carrying out the integration 
over the shaded area in Fig. \ref{fig_integ}.
Thus, this area is covered by ME if $p_{T{\rm min}} = {\hat p}_{T}$; 
{\it i.e.}, no radiation there.
PS has to cover the rest of the area.
We can easily see that the matching cannot be achieved by a simple choice 
of the PS energy scale.
A certain {\it rejection} method is necessary to apply, as is done in CKKW, 
when we use an ordinary $Q^{2}$-ordered PS.

We adopt the following method in the present study:
the PYTHIA PS is applied with the energy-scale choice of 
$\mu_{PS} = {\hat m}_{T}(W)$.
This can be done by explicitly setting the variable {\tt SCALUP} of 
the LHA common {\tt HEPEUP} to this value in the subroutine UPEVNT.
After the PS is added we investigate the parton information in the {\tt PYJETS} 
common of PYTHIA. 
The PS is re-applied to the same event if there is any radiation from the 
initial-state partons with $p_{T} \geq {\hat p}_{T}$.
This is an approximation since the examined $p_{T}$ does not directly 
correspond to the $p_{T}$ of each splitting.
However, this method can be applied without any modification to PYTHIA 
routines and, what is more, it provides us with a reasonably good results 
as will be shown in the following.
In any case, it is impossible to achieve a perfect matching using 
$Q^{2}$-ordered PSs.
There is no splitting to produce spacelike gluons in the $W$ + 1 jet ME, 
while in PS such a splitting may produce the highest $p_{T}$ jet 
although the probability must be very small.
This {\it rejection} does not significantly affect the event generation speed, 
because the PS simulation is fast and the average number of trials is only 
1.3 in the present condition.

\EPSFIGURE[t]{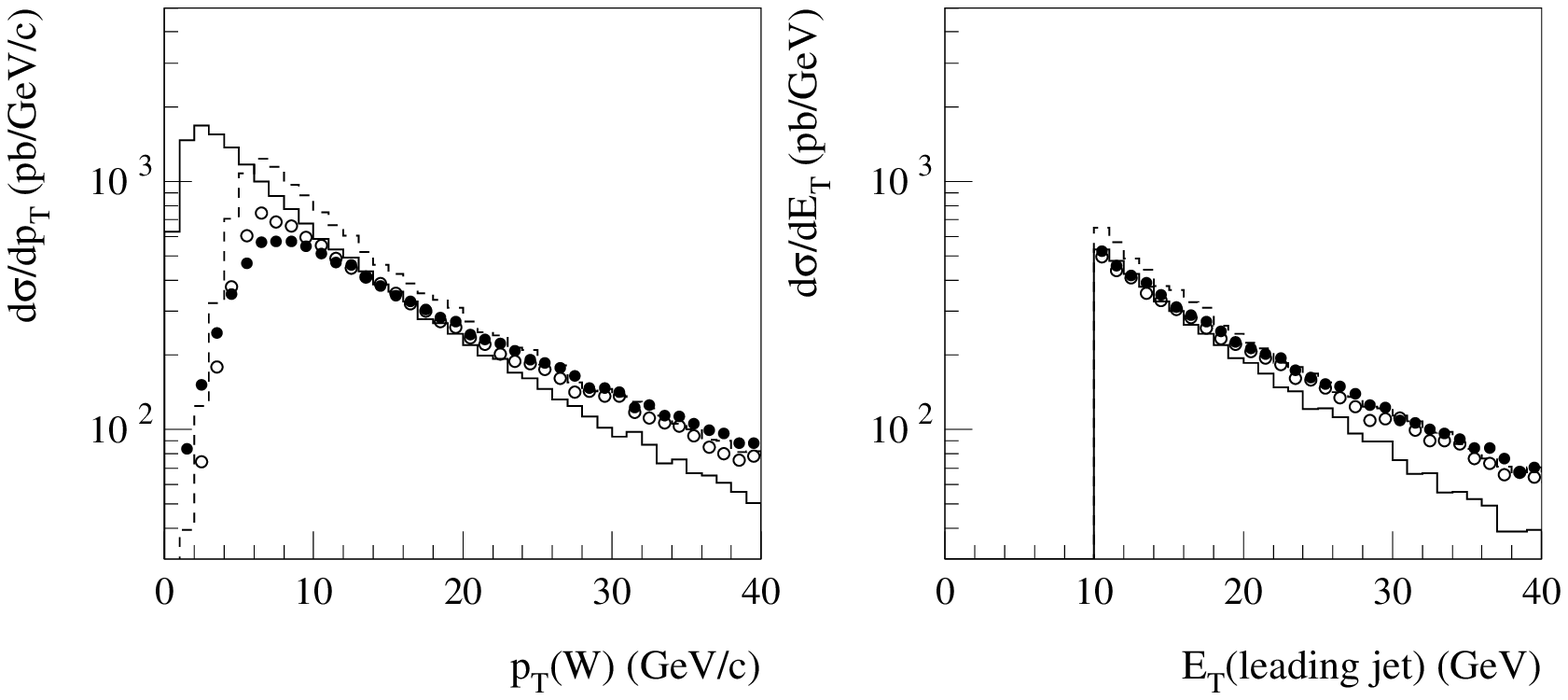,width=160mm}
{\label{fig_supp}
Suppressed $W$ + 1 jet simulation in the LHC condition.
A ${\hat p}_{T}$ cut of 5 GeV/$c$ is applied in the hard-interaction 
generation, and the jet reconstruction is applied with an $E_{T}$ threshold 
of 10 GeV.
Filled circles show the results of the simulation implementing 
the PS-ME matching, 
while open circles show those without matching but with 
$\mu_{PS} = {\hat p}_{T}$.
Dashed histograms are the naive simulation results for the energy-scale 
choice of $\mu = {\hat p}_{T}$.
Solid histograms show the results from the $W$ + 0 jet simulation with 
$\mu = m_{W}$.}

Figure \ref{fig_supp} shows the $p_{T}(W)$ and $E_{T}$(leading jet) 
distributions of the suppressed $W$ + 1 jet simulation.
Only low to medium $p_{T}$ and $E_{T}$ regions are shown in the figure.
The suppression is not significant in higher $p_{T}$ ($E_{T}$) regions.
The results from a naive simulation ($\mu = {\hat p}_{T}$) is overwritten 
with dashed histograms to show how the suppression works.
We can see a smooth matching with the $W$ + 0 jet simulation (histograms) 
at low $p_{T}$ ($E_{T}$) around 15 GeV.
This shows that our suppression method is reasonable.

Two results are plotted in the figure.
The filled circles show the results of the simulation implementing the 
above PS-ME matching method, 
while the open circles show those without matching.
The PS energy scale is set to ${\hat p}_{T}$ in the latter 
in order to avoid the overlap with ME.
The threshold behavior is more strongly smeared in the matched simulation.

\DOUBLEFIGURE[t]{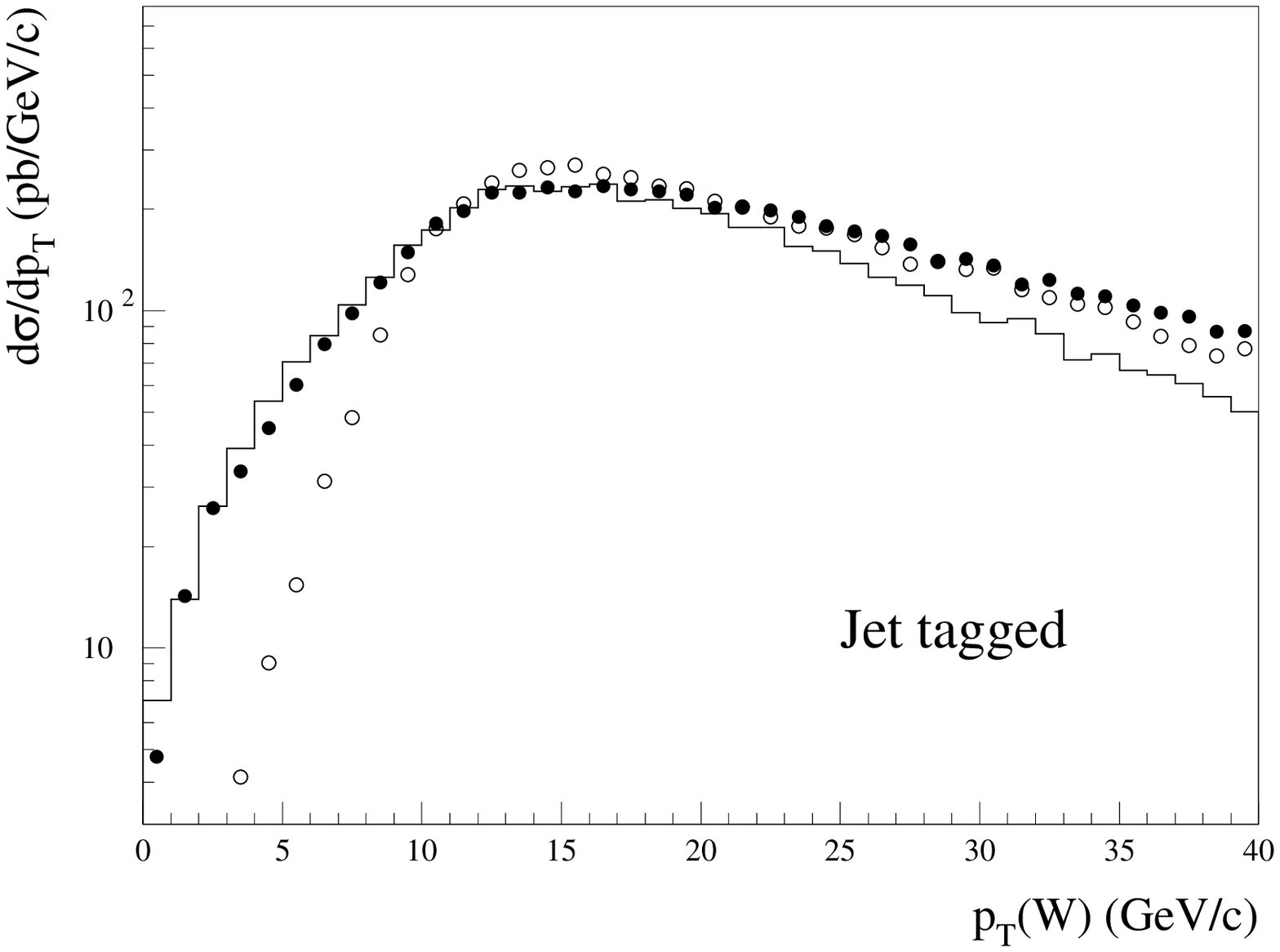,width=80mm}
{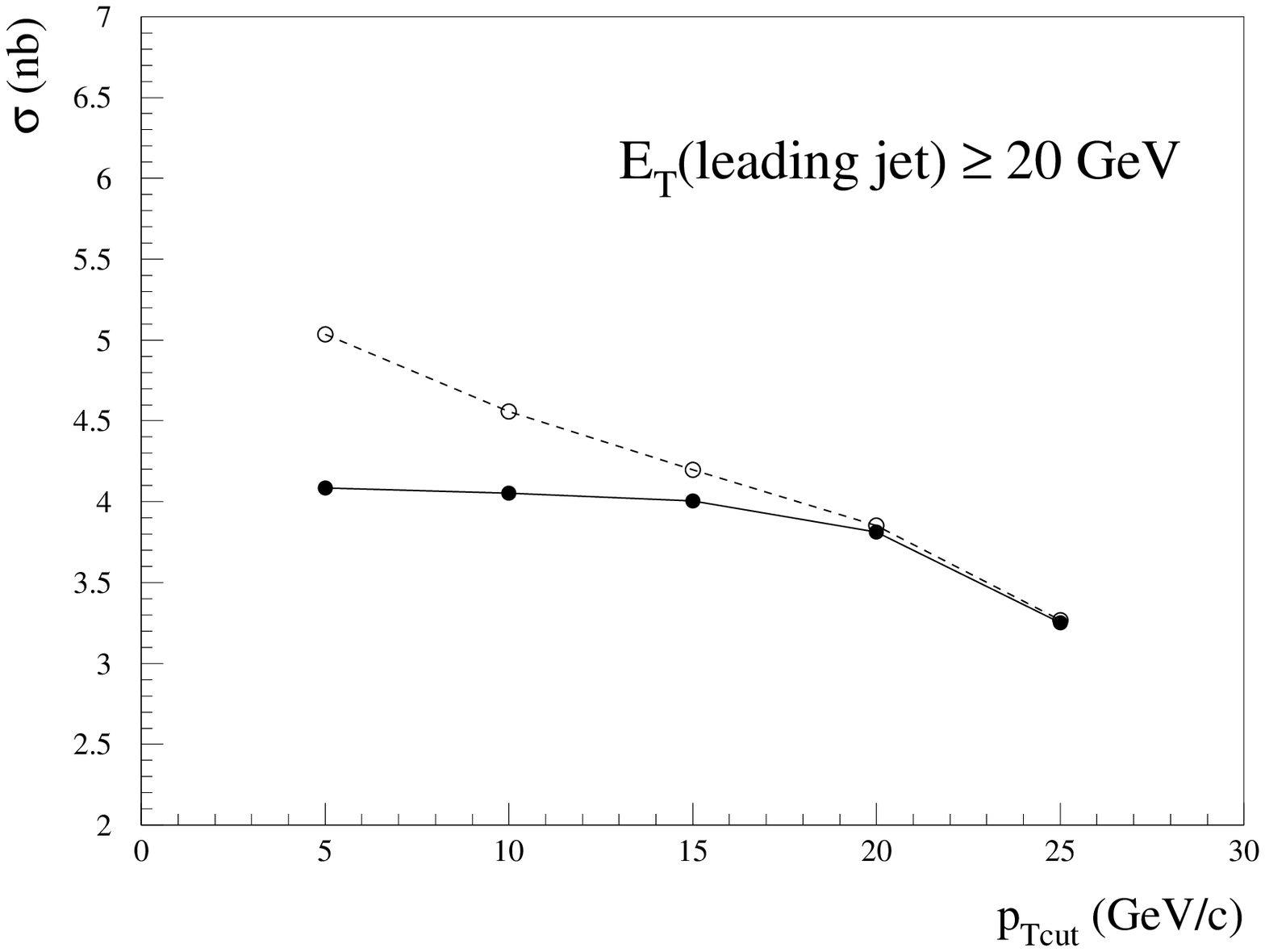,width=80mm}
{\label{fig_wpt_jtag}
The $p_{T}$ distribution of $W$ for jet-tagged events.
Filled circles show the result of the simulation to which the PS-ME matching 
method is applied, while open circles without the matching but with 
the PS energy-scale choice of $\mu_{PS} = {\hat p}_{T}$.
The histogram shows the result of the same analysis applied to the $W$ + 0 jet 
simulation.}
{\label{fig_ptcut}
Total cross section of jet-tagged events ($E_{T} \geq 20$ GeV) 
as a function of the ${\hat p}_{T}$-cut in the hard-interaction generation.
Filled circles show the results from the suppressed $W$ + 1 jet simulation 
with the PS-ME matching.
Open circles show those we obtain if we omit the {\it rejection} 
procedure in the matching.}


The difference between the matched and unmatched simulations is more clearly 
seen in Fig. \ref{fig_wpt_jtag}, 
where $p_{T}(W)$ is plotted for jet-tagged events.
The distribution below the $E_{T}$ threshold (10 GeV) shows the collective 
effect of the applied PS.
The result for the $W$ + 0 jet simulation obtained from the same analysis 
is overwritten in the figure.
The matched simulation is in good agreement with the $W$ + 0 
jet simulation below the threshold.
This means that our matching method is well performed; 
namely, the PS added to the $W$ + 1 jet simulation is quite similar to the 
non-leading PS in the $W$ + 0 jet simulation.
Since the leading behavior is already well matched, 
the suppressed $W$ + 1 jet simulation is now indistinguishable from
the $W$ + 0 jet simulation at low $p_{T}$s.


Figure \ref{fig_ptcut} shows the total cross section of jet-tagged events 
for a practical jet $E_{T}$ threshold (20 GeV), 
as a function of the ${\hat p}_{T}$ cut in GR@PPA.
Observable quantities should not depend on artificial conditions in the 
simulation such as the ${\hat p}_{T}$ cut.
The result shows that the jet-tagged cross section is stable against the 
variation of the ${\hat p}_{T}$ cut if it is set reasonably small.
The ${\hat p}_{T}$ cut of 10 GeV/$c$ looks sufficient in the present study,
though further smearing due to the hadronization and detector effects may 
alter it.

\section{$Z$ production at Tevatron Run I}

\EPSFIGURE[t]{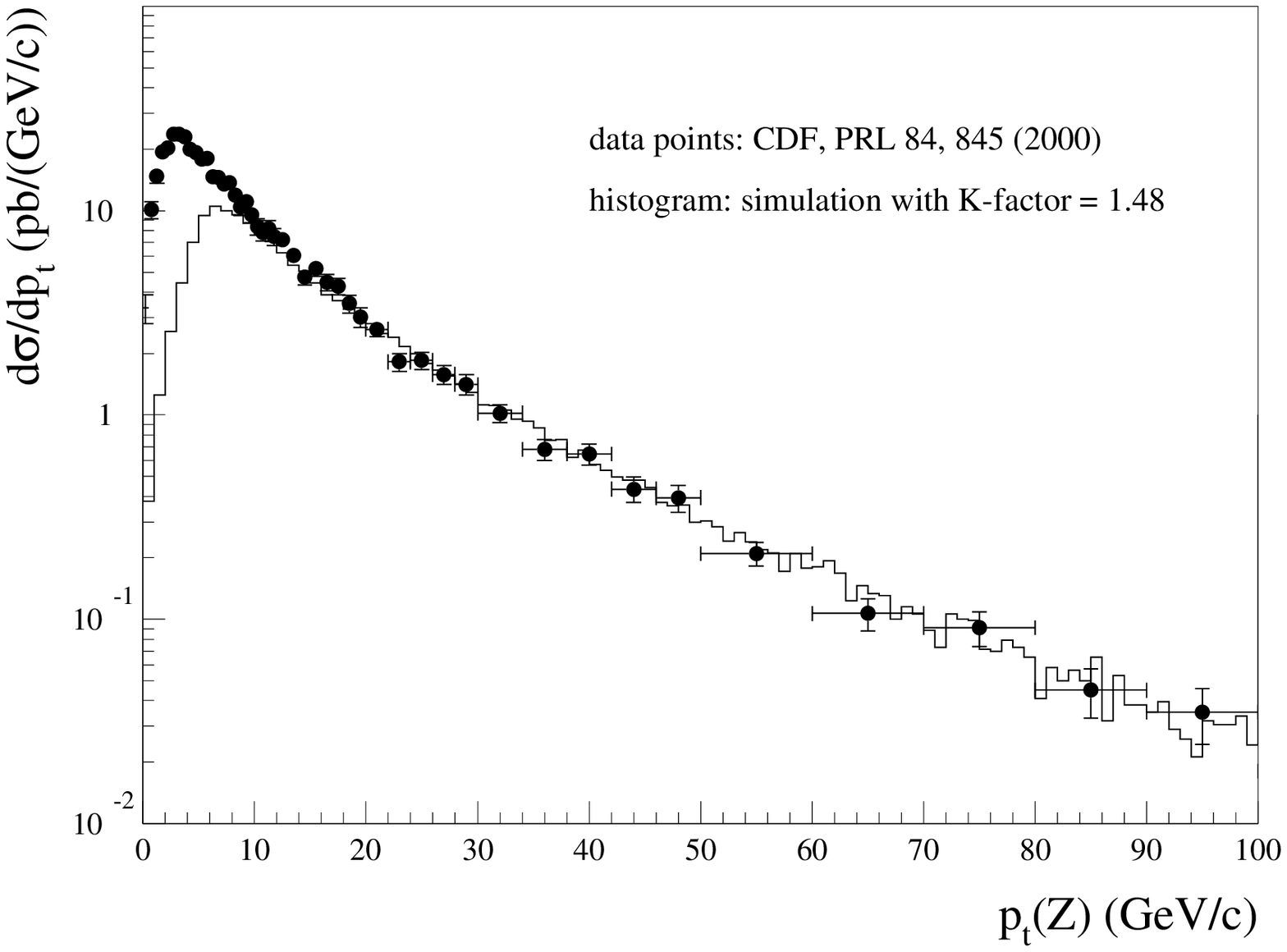,width=100mm}
{\label{fig_z_cdf}
Comparison with CDF data at Tevatron Run I for the $p_{T}$ spectrum of $Z$.
The histogram shows the simulation result for the $Z$ + 1 jet production 
with the ${\hat p}_{T}$ cut of 5 GeV/$c$.
}

In this section we apply our suppressed simulation to the $Z$ 
production at Tevatron for a comparison with experimental data.
The simulation for the $Z$ production ($66 < m(Z \rightarrow e^{+}e^{-}) 
< 116$ GeV/$c^{2}$) is compared with CDF data at Tevatron Run I \cite{cdf_z} 
in Fig. \ref{fig_z_cdf}.
The result from the suppressed $Z$ + 1 jet simulation for the Tevatron Run I 
condition (${\bar p}p$ collisions at 1.8 TeV in the cm energy) is shown 
with the histogram.
A ${\hat p}_{T}$ cut of 5 GeV/$c$ is applied again.
The simulation is essentially the same as that for the $W$ + 1 jet production, 
except for the overall normalization.
The simulation result is multiplied by a factor of 1.48, the ratio between 
the measured inclusive $Z$ production cross section and the corresponding 
tree-level prediction by the $Z$ + 0 jet generator in GR@PPA.
We can see a very good agreement from low ($\sim 15$ GeV/$c$) to very high 
($\sim 100$ GeV/$c$) $p_{T}$ regions.
It should be noted that we have obtained this simulation result without 
any tuning.

Unfortunately jet spectrum data are not available for the comparison.
There would be a difficulty in the presentation of experimental data 
since {\it jets} are experiment and analysis dependent.
By the way, there is no reason that experimental data much differ from our 
simulation, since the agreement is very good in the $p_{T}(Z)$ spectrum.
At the 0th order, the leading jet should be balanced with $Z$ in $p_{T}$.
Though higher order effects violate the balance, 
the dominant leading-order corrections are taken into account 
by applying a PS in our simulation.
Thus, significant deviation can emerge only if we examine a quantity relevant 
to the next-to-leading jet.
Note that the events generated by GR@PPA are passed to PYTHIA with the 
energy scale choice of $\mu_{PS} = {\hat m}_{T}(Z)$.
The final-state PS is implemented up to this scale 
since no rejection is applied to it.

\section{Conclusion}

We have introduced a simple method for simulating {\it inclusive} $W/Z$ + 
jet productions at hadron collisions.
The divergence of the cross section at low $p_{T}$ is suppressed 
by using the Sudakov form factor.
PS is added to the events generated by an ME-based event generator 
at the tree level with an appropriate consideration for matching.
The simulation can be done using existing tools, GR@PPA and PYTHIA.
Necessary modifications can be applied at the level where ordinary users are
allowed.

The results from our $W$ + 1 jet simulation show a smooth matching 
at low $p_{T}$s with the $W$ + 0 jet simulation, 
in which jet activities are totally generated by PS.
The simulation of $Z$ + 1 jet production in the Tevatron Run I condition 
well reproduces the CDF data for $p_{T}(Z)$ measurement.

The present study deals with the jet (parton) radiation from initial-state 
partons only.
An extension to the final state is necessary to apply our method to multi-jet 
productions.

The method applied in the present study is similar to the {\it inclusive} 
treatment for the highest jet-multiplicity events in CKKW \cite{sherpa}.
The success of our method suggests that the {\it inclusive} simulation of 
$n$-jet events in CKKW might be enough to obtain a reasonable 
inclusive $n$-jet sample, if the resolution scale can be lowered to the level 
of the ${\hat p}_{T}$ cut in the present study.
The contribution from other multiplicity events might become insignificant.

\acknowledgments

This work has been carried out as an activity of the NLO Working Group, 
a collaboration between the Japanese ATLAS group and the numerical analysis 
group (Minami-Tateya group) at KEK.
The author wants to acknowledge useful discussions with the members: 
J. Kodaira, J. Fujimoto, Y. Kurihara, T. Kaneko and T. Ishikawa of KEK, 
S. Tsuno of Okayama University, 
and K. Kato of Kogakuin University.

\listoftables		
\listoffigures		

\end{document}